# 二态隐马尔可夫过程熵率逼近序列的上下界

陈双平[①]  李军[①]  周密[①]

[①]（暨南大学电气信息学院 珠海 519070）

**摘要** 该文证明了一类二态隐马尔可夫过程熵率的逼近序列存在上下界，且其误差不大于一个因子小于 1 的几何级数，这说明逼近序列的收敛速度较快。该结论有助于理解一般隐马尔可夫过程熵率的收敛速度，并为部分隐马尔可夫过程熵率在任意误差范围内的估计提供了理论上的依据。

**关键词** 隐马尔可夫过程；熵率；上下界

**中图分类号** TN911.21, O236

# Low and Upper Bound of Approximate Sequence for the Entropy Rate of Binary Hidden Markov Processes

Chen Shuang-ping[①]  Li Jun[①]  Zhou Mi[①]

[①]（Electrical and Information College, Jinan University, Zhuhai ,519070,China）

**Abstract**  In the paper, the approximate sequence for entropy of some binary hidden Markov models has been found to have two bound sequences, the low bound sequence and the upper bound sequence. The error bias of the approximate sequence is bound by a geometric sequence with a scale factor less than 1 which decreases quickly to zero. It helps to understand the convergence of entropy rate of generic hidden Markov models, and it provides a theoretical base for estimating the entropy rate of some hidden Markov models at any accuracy.

**Key words** Hidden Markov Process；Entropy Rate；Low and Upper Bound

## 引言

  熵率在信息论里有着重要的理论和实际应用价值。对于离散无记忆信源和稳恒马尔可夫信源，其熵率有解析表达式或理论上的精确计算方法[1]，但是对于大多数离散信源来说，其熵率的计算都没有解析表达式,只能采取其他方法计算[1,2,3] ,或者分析其极限函数的性质[2]。尤其是对于稳恒隐马尔可夫信源，虽然我们业已证明了其熵率的存在性，但是计算其熵率需要计算一个表达式的极限[1]。这个极限是收敛的，但是这个极限的收敛速度不能计算。如果我们知道熵率表达式的紧上下界，这将有助于我们估计熵率。在给定精度的情况下，由上下界给出熵率的一个范围，只要上下界距离足够小，说明熵率的误差不会超出上下界的距离，此时我们就可以及时终止熵率的计算。

  我们在曾提出了一种思路，用迭代函数系统研究隐马尔可夫信源，并证明稳恒二态隐马

尔可夫信源熵率近似式的收敛速度是几何级数的。在 2006 年，我们应用该思路，提出了一个隐马尔可夫信源熵率的计算算法[4]。同时 Han 等采用迭代函数系统的思路证明了更一般情况下熵率的收敛性[5]。在本文中，我们提出并证明了两个上下界，它们为理解一般隐马尔可夫信源熵率的收敛速度提供了一种新的可能途径和解释，并为计算熵率的算法提供了可靠的误差估计方法。同时，不难将本文结论推广到更一般的情况。

基本概念

令 $X = \{X_k\}_{k \geq 1}$ 为二态字符集上的一阶静态马尔可夫过程，其转移概率 $P = \pi_{ab}$ 满足 $\pi_{ab} = P_X(X_k = b | X_{k-1} = a), a, b \in \{0,1\}$。另外有一个伯努利（Bernoulli）噪声过程（二态独立同分布）$E = \{E_k\}_{k \geq 1}$，独立于 $X$，满足 $P(E_i = 1) = \varepsilon$。因此，可以定义随机过程 $Z = \{Z_k\}_{k \geq 1}$，有 $Z_k = X_k \oplus E_k, k \geq 1$，这里 $\oplus$ 指模 2 加法（异或）。可以将 $Z$ 视为具有噪声 $E$ 的二态对称信道的输出，其输入为 $X$。$Z$ 完全由参数 $\pi_{01}$、$\pi_{10}$ 和 $\varepsilon$ 决定。随机过程 $Z$ 是一种最简单的隐马氏过程，简称为二态隐马尔可夫过程[2]。通常情况下我们认为噪声 $\varepsilon$ 较小，只考虑 $0 < \varepsilon, \pi_{01}, \pi_{10} < \frac{1}{2}$ 的情形。

我们将随机变量序列 $Z_i, ..., Z_j$ 记为 $Z_i^j$，大写字母代表随机变量序列，小写字母代表其一个实例。一般地，我们忽略某些地方的实例（如：$P(Z_i^j)$ 实际上意味着 $P(Z_i^j = z_i^j)$）。随机变量 $Z_i^j$ 的熵（在本文中我们采用信息熵，使用以 2 为底的对数，即 $\log x \equiv \log_2 x$）定义为[1,2]：

$H(Z_i^j) = E[-\log P(Z_i^j)].$

二态隐马尔可夫过程的熵率定义为[2]：

$h(Z) \equiv h(\pi_{01}, \pi_{10}, \varepsilon) = \lim_{n \to \infty} \frac{1}{n} H(Z_1^n) = \lim_{n \to \infty} \frac{1}{n} E[-\log P(Z_1^n)].$

由上可见计算二态隐马尔可夫过程的熵率，需要计算随机变量 $Z_1^n$ 的分布，这个分布有 $2^n$ 种取值的可能。下面我们来看如何计算 $P(Z_1^n)$。

定义 $g_0(x)$ 和 $g_1(x)$ 为：$g_0(x) = (1 - \pi_{01} - \pi_{10})(1 - 2\varepsilon)x + \pi_{10}(1 - \varepsilon) + (1 - \pi_{10})\varepsilon$, $g_1(x) = 1 - g_0(x)$. 由前后向算法[4]，知：$P(Z_{n+1} = 0 | Z_1^n) = g_0(P(X_n = 0 | Z_1^n))$, $P(Z_{n+1} = 1 | Z_1^n) = g_1(P(X_n = 0 | Z_1^n))$. 由此我们有[4]：

$P(Z_1^{n+1}) = P(Z_{n+1} | Z_1^n) P(Z_1^n)$
$= g_{z_{n+1}}(P(X_n = 0 | Z_1^n)) P(Z_1^n).$

定义 $f_0(x)$，$f_1(x)$ 为：

$f_0(x) = \frac{(1-\varepsilon)(x(1-\pi_{01}-\pi_{10})+\pi_{10})}{(1-\pi_{01}-\pi_{10})(1-2\varepsilon)x+\pi_{10}(1-\varepsilon)+(1-\pi_{10})\varepsilon}$, $f_1(x) = \frac{\varepsilon(x(1-\pi_{01}-\pi_{10})+\pi_{10})}{-(1-\pi_{01}-\pi_{10})(1-2\varepsilon)x+\pi_{10}\varepsilon+(1-\pi_{10})(1-\varepsilon)}$.

那么，我们有：

$P(X_n = 0 | Z_1^{n-1} 0) = f_0(P(X_{n-1} = 0 | Z_1^{n-1}))$, $P(X_n = 0 | Z_1^{n-1} 1) = f_1(P(X_{n-1} = 0 | Z_1^{n-1}))$. 综合这两式我们有：

$P(X_n = 0 | Z_1^{n-1} Z_{n+1}) = f_{z_{n+1}}(P(X_{n-1} = 0 | Z_1^{n-1})).$

定义两个函数的复合为：$f \circ g(x) \equiv f(g(x))$。定义 $F_{Z_1^n} \equiv f_{z_n} \circ f_{z_{n-1}} \circ \cdots \circ f_{z_1}$。那么我们有：$\forall n \in \aleph$，

$P(X_n = 0 | Z_1^n) = F_{Z_1^n}(P(X_0 = 0))$. 其中 $P(X_0 = 0)$ 代表二态隐马尔可夫过程的初始状态概率。

综合以上结果我们有 $P(Z_1^{n+1}) = g_{z_{n+1}}(P(X_n = 0 | Z_1^n))P(Z_1^n) = g_{z_{n+1}}(F_{Z_1^n}(P(X_0 = 0)))P(Z_1^n)$. 从而可以递推地计算得到 $P(Z_1^n), n = 1, 2, 3 \cdots$.

## 二态隐马尔可夫过程熵率的上下界

对于二态隐马尔可夫过程熵率也可以采用另一个公式[1] $h(Z) = \lim_{n \to \infty}(H(Z_1^{n+1}) - H(Z_1^n))$ 来计算。令 $h_b(x)$ 为定义于 [0,1] 区间的熵函数：$h_b(x) = -x \log x - (1-x) \log(1-x)$. 易见 $h_b(x)$ 在 $[0, \frac{1}{2}]$ 上单调增，在 $[\frac{1}{2}, 1]$ 上单调减，且关于 $x = \frac{1}{2}$ 对称 (即 $h_b(x) = h_b(1-x)$ )。同时注意到 $g_1(x) + g_0(x) = 1$ 对任意 $x$ 都成立，我们有

$$\begin{aligned}
H^{(n)} &= H(Z_1^{n+1}) - H(Z_1^n) \\
&= E[-\log P(Z_1^{n+1})] - E[-\log P(Z_1^n)] \\
&= E[-\log P(Z_1^n 0)] + E[-\log P(Z_1^n 1)] - E[-\log P(Z_1^n)] \\
&= \sum_{z_1^n} \left( -P(Z_1^n 0) \log P(Z_1^n 0) - P(Z_1^n 1) \log P(Z_1^n 1) \right) + \sum_{z_1^n} \left( P(Z_1^n) \log P(Z_1^n) \right) \\
&= \sum_{z_1^n} \left( -g_0(F_{Z_1^n}(P(X_0=0)))P(Z_1^n) \log\left(g_0(F_{Z_1^n}(P(X_0=0)))P(Z_1^n)\right) - g_1(F_{Z_1^n}(P(X_0=0)))P(Z_1^n) \log\left(g_1(F_{Z_1^n}(P(X_0=0)))P(Z_1^n)\right) \right) \\
&\quad + \sum_{z_1^n} \left( P(Z_1^n) \log P(Z_1^n) \right) \\
&= \sum_{z_1^n} \left( -g_0(F_{Z_1^n}(P(X_0=0)))P(Z_1^n) \log\left(g_0(F_{Z_1^n}(P(X_0=0)))\right) - g_1(F_{Z_1^n}(P(X_0=0)))P(Z_1^n) \log\left(g_1(F_{Z_1^n}(P(X_0=0)))\right) \right) \\
&\quad - \sum_{z_1^n} \left( g_0(F_{Z_1^n}(P(X_0=0)))P(Z_1^n) \log\left(P(Z_1^n)\right) + g_1(F_{Z_1^n}(P(X_0=0)))P(Z_1^n) \log\left(P(Z_1^n)\right) \right) + \sum_{z_1^n} \left( P(Z_1^n) \log P(Z_1^n) \right) \\
&= \sum_{z_1^n} \left( -g_0(F_{Z_1^n}(P(X_0=0)))P(Z_1^n) \log\left(g_0(F_{Z_1^n}(P(X_0=0)))\right) - g_1(F_{Z_1^n}(P(X_0=0)))P(Z_1^n) \log\left(g_1(F_{Z_1^n}(P(X_0=0)))\right) \right) \\
&\quad - \sum_{z_1^n} \left( P(Z_1^n) \log P(Z_1^n) \right) + \sum_{z_1^n} \left( P(Z_1^n) \log P(Z_1^n) \right) \\
&= \sum_{z_1^n} \left( -g_0(F_{Z_1^n}(P(X_0=0)))P(Z_1^n) \log\left(g_0(F_{Z_1^n}(P(X_0=0)))\right) - g_1(F_{Z_1^n}(P(X_0=0)))P(Z_1^n) \log\left(g_1(F_{Z_1^n}(P(X_0=0)))\right) \right) \\
&= \sum_{z_1^n} P(Z_1^n) h_b\left(g_0(F_{Z_1^n}(P(X_0=0)))\right)
\end{aligned}$$

对于 $f_0(x)$ 和 $f_1(x)$，不难证明它们都是 [0,1] 区间的单调增函数[4]，因此 $F_{Z_1^n}(x)$ 也是在 [0,1] 区间上的单调增函数。显然对于 $f_0(x)$ 和 $f_1(x)$，有 $f_0([0,1]) \subset [0,1], f_1([0,1]) \subset [0,1]$。

因为 $H^{(n)} = \sum_{z_1^n} P(Z_1^n) h_b\left(g_0(F_{Z_1^n}(P(X_0=0)))\right)$ 取决于 $P(X_0 = 0)$ 的取值，实际上对于本文给出的二态隐马尔可夫过程，其底层的马尔可夫模型是稳恒遍历的（在本文中 0，1 状态是对称的），因此在计算时取稳恒态 $P(X_0 = 0) = P(X_t = 0) = \frac{1}{2}$，这样由稳恒性，我们可以得出对任意正整数 $t$，$P(Z_1^n) = P(Z_{1+t}^{n+t})$。

我们引入 $L^{(n)}, U^{(n)}$ 如下：

$$L^{(n)} = \sum_{z_1^n} P(Z_1^n) \min_{x \in [0,1]} \{h_b(g_0(F_{Z_1^n}(x)))\}$$
$$<= \sum_{z_1^n} P(Z_1^n) h_b\left(g_0(F_{Z_1^n}(P(X_0=0)))\right) = \sum_{z_1^n} P(Z_1^n) h_b\left(g_0(F_{Z_1^n}(\frac{1}{2})))\right) = H^{(n)}$$
$$<= U^{(n)} = \sum_{z_1^n} P(Z_1^n) \max_{x \in [0,1]} \{h_b(g_0(F_{Z_1^n}(x)))\}$$

我们已经知道 $H^{(n)}$ 是收敛的，因此易见 $L^{(n)}$ 有上界且 $U^{(n)}$ 有下界，下面我们只需要证明 $L^{(n)}$ 和 $U^{(n)}$ 收敛。实际上我们只需证明 $U^{(n)} - L^{(n)}$ 趋于零，如果它成立，由 $|H^{(n)} - L^{(n)}| \le U^{(n)} - L^{(n)}, |H^{(n)} - U^{(n)}| \le U^{(n)} - L^{(n)}$ 和 $H^{(n)}$ 收敛可得：$L^{(n)}$，$U^{(n)}$ 和 $H^{(n)}$ 都收敛于同一个极限就是熵率 $h(Z)$。

如果存在合适的参数 $\pi_{01}$、$\pi_{10}$ 和 $\varepsilon$（比如 $\pi_{01} = \pi_{10} = 0.1, \varepsilon = 0.01$），使得 $\exists \delta > 0, \forall x \in [0,1], |f_0'(x)| < \delta < 1, |f_1'(x)| < \delta < 1$。那么，存在 $\eta_1, \eta_2, \cdots \eta_{n-1} \in [0,1]$，使得

$$\left|F_{Z_1^n}'(x)\right| \equiv \left|\left(f_{z_n} \circ f_{z_{n-1}} \circ \cdots f_{z_1}\right)'(x)\right|$$
$$= \left|f_{z_n}'(\eta_1)\right|\left|f_{z_{n-1}}'(\eta_2)\right|\cdots\left|f_{z_1}'(x)\right|$$
$$< \delta^n$$

定义函数 $h_{g0}(x) \equiv h_b(g_0(x))$，因 $g_0(x)$ 在 $[0,1]$ 区间连续可导，$g_0([0,1]) \subset (0,1)$，$h_b(x)$ 在 $(0,1)$ 内的闭区间连续可导，因此 $h_{g0}(x)$ 在在 $[0,1]$ 区间连续可导，且其导数还是连续的，因此其导数有界，设 $|h_{g0}'(x)| < M$。我们有：

$$U^{(n)} - L^{(n)} = \sum_{z_1^n} P(Z_1^n) \left( \max_{x \in [0,1]} \left\{h_b\left(g_0\left(F_{Z_1^n}(x)\right)\right)\right\} - \min_{x \in [0,1]} \left\{h_b\left(g_0\left(F_{Z_1^n}(y)\right)\right)\right\} \right)$$
$$= \sum_{z_1^n} P(Z_1^n) \left( \max_{y \in \left[F_{Z_1^n}(0), F_{Z_1^n}(1)\right]} \left\{h_b\left(g_0(y)\right)\right\} - \min_{y \in \left[F_{Z_1^n}(0), F_{Z_1^n}(1)\right]} \left\{h_b\left(g_0(y)\right)\right\} \right)$$
$$= \sum_{z_1^n} P(Z_1^n) \left( \max_{y \in \left[F_{Z_1^n}(0), F_{Z_1^n}(1)\right]} \left\{h_{g0}(y)\right\} - \min_{y \in \left[F_{Z_1^n}(0), F_{Z_1^n}(1)\right]} \left\{h_{g0}(y)\right\} \right)$$
$$= \sum_{z_1^n} P(Z_1^n) \left|h_{g0}'(\eta)\right|\left|F_{Z_1^n}(1) - F_{Z_1^n}(0)\right|$$
$$< \sum_{z_1^n} P(Z_1^n) M \left|F_{Z_1^n}(1) - F_{Z_1^n}(0)\right|$$
$$\le \sum_{z_1^n} P(Z_1^n) M \left|F_{Z_1^n}'(\zeta)\right||1-0|$$
$$< \sum_{z_1^n} P(Z_1^n) M \delta^n$$
$$= M \delta^n$$

因为 $M$ 为常数，而 $\delta < 1$，因此 $U^{(n)} - L^{(n)}$ 收敛于 0。

由本文考虑的隐马尔可夫过程是稳恒遍历的，因此任意正整数 $t$，$P(Z_1^n) = P(Z_{1+t}^{n+t})$。结合上面 $f_0([0,1]) \subset [0,1], f_1([0,1]) \subset [0,1]$，我们可以得到：

$$\begin{aligned}
L^{(n+1)} &= \sum_{z_1^{n+1}} P(Z_1^{n+1}) \min_{x \in [0,1]} \left\{ h_b \left( g_0 \left( F_{Z_1^{n+1}}(x) \right) \right) \right\} \\
&= \sum_{z_1 z_2^{n+1}} P(Z_1 Z_2^{n+1}) \min_{x \in [0,1]} \left\{ h_b \left( g_0 \left( F_{Z_1 Z_2^{n+1}}(x) \right) \right) \right\} \\
&= \sum_{0 z_2^{n+1}} P(0 Z_2^{n+1}) \min_{x \in [0,1]} \left\{ h_b \left( g_0 \left( F_{0 Z_2^{n+1}}(x) \right) \right) \right\} + \sum_{1 z_2^{n+1}} P(1 Z_2^{n+1}) \min_{x \in [0,1]} \left\{ h_b \left( g_0 \left( F_{1 Z_2^{n+1}}(x) \right) \right) \right\} \\
&= \sum_{z_2^{n+1}} P(0 Z_2^{n+1}) \min_{x \in [0,1]} \left\{ h_b \left( g_0 \left( F_{Z_2^{n+1}}(f_0(x)) \right) \right) \right\} + \sum_{z_2^{n+1}} P(1 Z_2^{n+1}) \min_{x \in [0,1]} \left\{ h_b \left( g_0 \left( F_{Z_2^{n+1}}(f_1(x)) \right) \right) \right\} \\
&= \sum_{z_1^n} P(0 Z_1^n) \min_{x \in [0,1]} \left\{ h_b \left( g_0 \left( F_{Z_1^n}(f_0(x)) \right) \right) \right\} + \sum_{z_1^n} P(1 Z_1^n) \min_{x \in [0,1]} \left\{ h_b \left( g_0 \left( F_{Z_1^n}(f_1(x)) \right) \right) \right\} \\
&= \sum_{z_1^n} P(0 Z_1^n) \min_{y \in [f_0(0), f_0(1)]} \left\{ h_b \left( g_0 \left( F_{Z_1^n}(y) \right) \right) \right\} + \sum_{z_1^n} P(1 Z_1^n) \min_{y \in [f_1(0), f_1(1)]} \left\{ h_b \left( g_0 \left( F_{Z_1^n}(y) \right) \right) \right\} \\
&\geq \sum_{z_1^n} P(0 Z_1^n) \min_{y \in [0,1]} \left\{ h_b \left( g_0 \left( F_{Z_1^n}(y) \right) \right) \right\} + \sum_{z_1^n} P(1 Z_1^n) \min_{y \in [0,1]} \left\{ h_b \left( g_0 \left( F_{Z_1^n}(y) \right) \right) \right\} \\
&= \sum_{z_1^n} \left( P(0 Z_1^n) + \sum P(1 Z_1^n) \right) \min_{y \in [0,1]} \left\{ h_b \left( g_0 \left( F_{Z_1^n}(y) \right) \right) \right\} \\
&= \sum_{z_1^n} P(Z_1^n) \min_{y \in [0,1]} \left\{ h_b \left( g_0 \left( F_{Z_1^n}(y) \right) \right) \right\} \\
&= L^{(n)}
\end{aligned}$$

由此可见 $L^{(n)}$ 是单调不减序列。同理可证 $U^{(n)}$ 是单调不增序列，因此 $L^{(n)} \leq L^{(n+1)} \leq \lim_{n \to \infty} L^{(n)} = h(Z) = \lim_{n \to \infty} H^{(n)} = \lim_{n \to \infty} U^{(n)} \leq U^{(n+1)} \leq U^{(n)}$，这也就是说 $H^{(n)}, h(Z)$ 包含于区间 $\left[ L^{(n)}, U^{(n)} \right]$

因此 $|H^{(n)} - h(Z)| \leq U^{(n)} - L^{(n)} \leq M \delta^n$，也就是说熵率逼近序列 $H^{(n)}$ 的收敛速度最少是几何级数级。

结论

本文证明了对一类特殊的二态隐马尔可夫过程，只要其参数 $\pi_{01}$、$\pi_{10}$ 和 $\varepsilon$ 合适，使得 $\forall x \in [0,1], |f_0'(x)| < \delta < 1, |f_1'(x)| < \delta < 1$。则二态隐马尔可夫过程的熵率 $h(Z)$ 存在不减的下序列界 $L^{(n)}$ 和不增的上界序列 $U^{(n)}$，满足：

$$L^{(n)} = \sum_{z_1^n} P(Z_1^n) \min_{x \in [0,1]} \{ h_b(g_0(F_{Z_1^n}(x))) \} \leq h(Z) \leq U^{(n)} = \sum_{z_1^n} P(Z_1^n) \max_{x \in [0,1]} \{ h_b(g_0(F_{Z_1^n}(x))) \}.$$

也就是说 $h(Z)$ 是区间 $\left[ L^{(n)}, U^{(n)} \right]$ 中的一个数，这一结论为估计熵率提供了可能。并且我们还证明了 $|H^{(n)} - h(Z)| \leq U^{(n)} - L^{(n)} \leq M \delta^n$，从而得出熵率逼近序列 $H^{(n)}$ 的误差最少是几何级数。

文中涉及的条件虽然很特殊，但是在很多情况下是满足的[6]。同时本文这些证明为一般情形的证明提供了可供参考的途径，本文中的上下界也为熵率的计算提供了中止的条件，可以用本文的结论求出一定精度的熵率结果。从这个意义上来讲，本文的结论和方法具有重要的价值。

作者简介：
陈双平：男，1976 年生，博士，研究兴趣为复杂性科学，生物信息学。
李军：男，1973 年生，博士/副教授，研究兴趣为智能信息处理。
周密：男，1977 年生，博士，研究兴趣为数据挖掘/智能信息处理。